\documentclass[10pt,conference]{IEEEtran}

\usepackage[normalem]{ulem}
\usepackage{dblfloatfix}
\usepackage{mdframed}
\usepackage{verbatimbox}
\usepackage{fancyvrb}
\usepackage{paralist}
\usepackage{etex}
\usepackage{pstricks}
\usepackage{tikz}
\usetikzlibrary{positioning}
\usepackage{balance}
\usetikzlibrary{shadows,arrows}
\usetikzlibrary{patterns,calc}
\usepackage{calc}
\usepackage{booktabs}
\usepackage{float}
\usepackage{multicol}
\usepackage{graphicx,subcaption}
\usepackage{multirow}
\usepackage{syntax}
\usepackage{color, colortbl}
\usepackage{enumerate}
\usepackage{wrapfig}
\usepackage{caption,subcaption}
\usepackage{hyperref}
\usepackage{listings}
\usepackage{pgfplots}
\usepackage{tabu}
\usepackage{amsmath}
\usepackage{amsfonts}
\usepackage{enumitem}
\usepackage{subcaption}
\usepackage{capt-of}

\usepackage[noend,linesnumbered,ruled]{algorithm2e}

\usepackage[noend]{algpseudocode}
\SetAlFnt{\small}
\hypersetup{
     colorlinks = true,
     citecolor = blue
}
\usepackage{breakurl}
\usepackage{url}
\usepackage{array}
\newcolumntype{P}[1]{>{\centering\arraybackslash}p{#1}}

\definecolor{coolblack}{rgb}{0.0, 0.18, 0.39}
\definecolor{carnelian}{rgb}{0.7, 0.11, 0.11}
\definecolor{antiquefuchsia}{rgb}{0.57, 0.36, 0.51}
\definecolor{airforceblue}{rgb}{0.36, 0.54, 0.66}
\definecolor{ferngreen}{rgb}{0.31, 0.47, 0.26}
\definecolor{darkpastelred}{rgb}{0.76, 0.23, 0.13}
\definecolor{bluebell}{rgb}{0.64, 0.64, 0.82}
\definecolor{bluegray}{rgb}{0.4, 0.6, 0.8}
\definecolor{bluepigment}{rgb}{0.2, 0.2, 0.6}
\definecolor{cerulean}{rgb}{0.0, 0.48, 0.65}
\definecolor{grannysmithapple}{rgb}{0.66, 0.89, 0.63}
\definecolor{junglegreen}{rgb}{0.16, 0.67, 0.53}
\definecolor{lightgreen}{rgb}{0.56, 0.93, 0.56}
\definecolor{salmon}{rgb}{1.0, 0.55, 0.41}
\definecolor{palemagenta}{rgb}{0.98, 0.52, 0.9}
\definecolor{rose}{rgb}{1.0, 0.0, 0.5}
\definecolor{icterine}{rgb}{0.99, 0.97, 0.37}
\definecolor{laserlemon}{rgb}{1.0, 1.0, 0.13}
\definecolor{darkbyzantium}{rgb}{0.36, 0.22, 0.33}
\definecolor{regalia}{rgb}{0.32, 0.18, 0.5}
\definecolor{asparagus}{rgb}{0.53, 0.66, 0.42}
\definecolor{cadmiumgreen}{rgb}{0.0, 0.42, 0.24}
\definecolor{columbiablue}{rgb}{0.61, 0.87, 1.0}
\definecolor{alizarin}{rgb}{0.82, 0.1, 0.26}
\definecolor{britishracinggreen}{rgb}{0.0, 0.26, 0.15}
\definecolor{ferrarired}{rgb}{1.0, 0.11, 0.0}
\definecolor{prussianblue}{rgb}{0.0, 0.19, 0.33}
\definecolor{uclagold}{rgb}{1.0, 0.7, 0.0}
\definecolor{tablecolor}{rgb}{0.2, 0.2, 0.6}

\usepackage[capitalize]{cleveref}

\begin{document}
\title{Load-Balanced Sparse MTTKRP on GPUs}

\author{put the authors}
\author{
    \IEEEauthorblockN{Israt Nisa\IEEEauthorrefmark{2},
		      Jiajia Li\IEEEauthorrefmark{3},
                      Aravind Sukumaran-Rajam\IEEEauthorrefmark{2},
                      Richard Vuduc\IEEEauthorrefmark{4},
                      P. Sadayappan\IEEEauthorrefmark{2}
    }

\IEEEauthorblockA{\IEEEauthorrefmark{2}
      {The Ohio State University, Columbus OH, USA}}
\IEEEauthorblockA{\IEEEauthorrefmark{3}
      {Pacific Northwest National Laboratory, Richland WA, USA}}
\IEEEauthorblockA{\IEEEauthorrefmark{4}
      {Georgia Institute of Technology, Atlanta GA, USA}}

Email: \{nisa.1, sukumaranrajam.1, sadayappan.1\}@osu.edu, jiajia.li@pnnl.gov,  richie@cc.gatech.edu
}


\maketitle

\begin{abstract}
Sparse matricized tensor times Khatri-Rao product (MTTKRP) is one of the most computationally expensive kernels in sparse tensor computations. This work focuses on optimizing the MTTKRP operation on GPUs, addressing both performance and storage requirements.  We begin by identifying the performance bottlenecks in directly extending the state-of-the-art CSF (compressed sparse fiber) format from CPUs to GPUs.  A significant challenge with GPUs compared to multicore CPUs is that of utilizing the much greater degree of parallelism in a load-balanced fashion for irregular computations like sparse MTTKRP.  To address this issue, we develop a new storage-efficient representation for tensors that enables high-performance, load-balanced execution of MTTKRP on GPUs.  A GPU implementation of sparse MTTKRP using the new sparse tensor representation is shown to outperform all currently known parallel sparse CPU and GPU MTTKRP implementations.

\end{abstract}

\section{Introduction}
Tensors are used to represent high dimensional data. For example, 
the attributes of an email conversation (subject, author and time) can 
be represented by the use of a tensor or a 3-way array. 
Multidimensional data is widely used in healthcare analytics 
\cite{ho2014limestone}, \cite{ho2014marble}, telecommunications 
\cite{sun2006window}, \cite{sun2006beyond}, machine learning 
\cite{abadi2016tensorflow}, text analysis \cite{bader2008discussion} 
among many other applications. Decomposition of such tensors was first utilized in the field of 
psychometrics in 1970 \cite{harshman1970foundations}, followed by the 
field of chemometrics eleven years later \cite{appellof1981strategies}.
 Over the last few years, such decomposition techniques have gained 
 significant popularity because of their applicability in recommender 
 systems, conversation detection etc.

Canonical Polyadic Decomposition (CPD) is one of the most common tensor 
factorization techniques, applicable to both dense and sparse 
tensors  \cite{kolda2009tensor}. MTTKRP (multiplying a sparse matricized
 tensor by Khatri-Rao product), a key kernel, is a common bottleneck for
 CPD. The remaining kernels for CPD are highly optimized in BLAS libraries,
  making optimization of MTTKRP the key area of focus for many researchers. 

Real world sparse tensors are extremely large, often follow power-law 
distribution, and are extremely sparse. 
Several efforts have addressed the
 parallelization of MTTKRP -  DFacTo \cite{choi2014dfacto}, GigaTensor 
 \cite{kang2012gigatensor}, SPLATT \cite{smith2015splatt}, HiCOO \cite{hicoo-li}, CSTF \cite{Blanco:2018}, and others \cite{choi2018blocking,Baskaran:2017} 
 for CPUs, FCOO \cite{liu2017unified}, ParTI \cite{parti} 
 for GPUs, and one \cite{smith2017knl} for the Intel Xeon Phi. A simple and popular 
 format of storing tensors is COO (coordinate), where indices for all the 
 dimensions are stored along with each non zero value. The state-of-the-art 
 CPU implementation of sparse MTTKRP, SPLATT, stores tensors in the CSF (compressed sparse fiber) format, 
a hierarchical 
 storage format. CSF for tensors is similar to CSR (compressed sparse row) for 
 matrices. To avoid repetitive row entries, CSR stores a pointer to the start 
 of a row. However, for hyper-sparse matrices, where a significant number of rows could be empty,
 DCSR (doubly compressed sparse row) 
 is a more efficient choice. In DCSR, row indices are also stored along with 
 row pointers, but 
 only for the non-empty rows. CSF is an extension of DCSR to tensors.

The CSF format provides many benefits for sparse MTTKRP compared to the COO
format, including reduced volume of data movement from main memory and
reduced floating-point operation count (as explained later in the paper)
and avoidance/reduction of expensive atomic operations. A significant
challenge with the CSF format is load balancing, when the tensor structure
exhibits very high variance in the distribution of nonzeros across fibers/slices.
Fibers and slices are two kinds of subsets of a tensor, corresponding to 1D vectors and 2D matrices, respectively.
Some recent efforts have addressed the problem of achieving the conflicting
demands of good load-balance and efficient data access for multi-core CPUs 
\cite{choi2018blocking,hicoo-li}.
But the load-balancing problem is especially acute for a GPU implementation
of sparse MTTKRP due to the significantly higher degree of thread-level parallelism
that must be utilized, when compared to multicore CPUs.

In this paper, we first undertake performance analysis 
across a range of datasets
using a GPU-CSF implementation of sparse MTTKRP,
which provides insights into a number of
load-balancing and data access issues: 
\begin{compactitem}
\item Significant inter-warp load imbalance and low GPU occupancy when the standard deviation of
fiber length in a mode is high;
\item Significant inter-thread-block load imbalance for the short-fiber mode when 
the standard deviation of the outer-most dimension's slice volume is high;
\item High data access overhead for ultra-sparse tensors where average slice volume
is extremely low, close to 1.
\end{compactitem}

Based on these insights, we develop new sparse tensor formats to address two
complementary aspects:
1) The fundamental difference between the parallel structure of threads
in GPUs versus multicore CPUs, necessitating attention to load-balancing at
both levels (between warps in a thread block and thread blocks in a grid);  
2) The diversity of the nonzero distribution patterns in a sparse tensor, 
with different representations being beneficial for ultra-sparse versus
moderately sparse regions of a tensor.
In summary, the key contributions in this work are:
\begin{itemize}
\item We propose a Balanced-CSF (B-CSF) format for GPUs, as an extension of the state-of-the-art CSF format for multicore processors. B-CSF uses a combination of fiber splitting and slice splitting to enable the good load-balancing on GPUs.
\item We construct a Hybrid Balanced-CSF (HB-CSF) format by partitioning of the nonzeros of a sparse tensor into three groups with different representations -- B-CSF, CSL, and COO.
\item By using the HB-CSF format, we implement a new sparse GPU-MTTKRP algorithm that achieves
significant speedup over existing implementations on CPUs and GPUs. 

\end{itemize}

\section{Background}
\label{sec:background}
We begin by summarizing the basic tensor notations used in the paper,
and then briefly describe the basics of the CPD and MTTKRP computation. A more
detailed description of tensor decomposition methods developed over the last
few decades, along with their applications, can be found in the survey by Kolda
and Bader \cite{kolda2009tensor}.

\subsection{Tensor Notation}
We follow the definitions and symbols used by Kolda and Sun in
\cite{kolda2008scalable}. A tensor can be visualized as a multi-dimensional
array. The order of a tensor refers to the number of dimensions or modes; an
order $N$ tensor has $N$ dimensions or modes. Tensors of order $N\geq3$ are
represented by capital caligraphic letters (e.g., $\mathcal{X}$).  Tensors of
order 2, i.e, 2D matrices, are denoted by capital letters (e.g., $A$). Tensors of
order 1, i.e, vectors, are denoted by lowercase letters (e.g., $a$). For the rest
of the paper, unless otherwise noted, a third-order (3D) tensor $\mathcal{X}$ is of
dimensions $I \times J \times K$. An element at $(i,j,k)$ position of
tensor $\mathcal{X}$ is denoted by $\mathcal{X}_{ijk}$, an element at $(i,j)$
position of matrix $A$, by $A_{ij}$, and an element at $i$  position of vector $a$,
by $a_i$.

A \textit{fiber} is a vector constructed from fixing all but one index, and is
denoted as $\mathcal{X}_{:,j,k}$. Similarly, a \textit{slice}, denoted as
$\mathcal{X}_{:,:,k}$, is a matrix obtained by fixing all but two indices. The
other symbols commonly used in the paper are summarized in Table
\ref{tbl:notation}.

\begin{table}[htb]
\scriptsize
\center
\begin{tabular}{c|l}
\hline
Notation & Description            \\ \hline
$\mathcal{X}$        & Tensor        \\ 
$N$        & Tensor order \\ 
$R$        & Tensor rank        \\ 
$M$        & Number of nonzeros in $\mathcal{X}$  \\ 
$S$        & Number of slices in $\mathcal{X}$     \\ 
$F$        & Number of fibers in $\mathcal{X}$     \\ 
$\mathcal{X}_{ijk}$     & Element at $(i,j,k)$ of $\mathcal{X}$  \\ 
$\mathcal{X}_{:,j,k}$     & Column fiber of $\mathcal{X}$       \\ 
$\mathcal{X}_{i,:,:}$     & Horizontal slice of $\mathcal{X}$   \\ \hline
$A$        & Matrix                 \\ 
$a$        & Vector                 \\ 
$A_{ij}$      & Element at index $(i,j)$ of A    \\ 
$a_i$       & Element at index $i$ of a      \\ 
\hline
\end{tabular}
\caption{Tensor notations}
\label{tbl:notation}
\end{table}

\subsection{CANDECOMP/PARAFAC (CP)}

CPD is a technique frequently used to decompose both sparse and dense tensors.
Its function is similar to Singular Value Decomposition (SVD), which discovers
latent features of data in matrices. CPD decomposes a tensor into a sum of
component rank-one tensors. The tensor $\mathcal{X}$ can then be mathematically
approximated as, 
\begin{align}
 \mathcal{X} \approx \sum_{r=1}^{R} \lambda_r a_r \circ b_r \circ c_r = [[ \lambda; A, B, C ]]
 \label{eq:sumVec}
\end{align}

where $a_r \in \mathbb{R}^{I}$, $b_r \in \mathbb{R}^{J}$ and $c_r \in
\mathbb{R}^{K}$. $A \in \mathbb{R}^{I \times R}$, $B \in \mathbb{R}^{J \times
R}$ and $C \in \mathbb{R}^{K \times R}$ are the factor matrices, constructed
using the corresponding rank-one tensors, for example, $ A = [a_1 \  a_2 \   ...\
a_R ] $.  $\lambda \in  \mathbb{R}^{R}$ stores the weights of the normalized
columns of the matrices.


Among many CP decomposition algorithms, the alternating least squares (ALS)
method \cite{carroll1970analysis}, \cite{harshman1970foundations} is one of the
most popular, applicable to both dense and sparse tensors. To approximate
$\mathcal{X}$, we look for the optimal solution of $\min \ || \mathcal{X} -
\widetilde{\mathcal{X}} || $, where $\widetilde{\mathcal{X}}$ is the approximation
computed using Equation \ref{eq:sumVec}. In ALS, one factor
matrix is updated at a time, while the rest are fixed. For example, the 
following equation solves $A$, while $B$ and $C$ are kept fixed. This minimization 
problem is reformulated by solving the following Frobenius Norm equation:
\begin{align}
 \parallel X_{(1)} -  \widetilde{A}(C \odot B)^T  \parallel_F
 \label{eq:min}
\end{align}

Similarly, $B$ and $C$ are solved by fixing $\{A,C\}$ and $\{A,B\}$ respectively.
Equation (\ref{eq:min}) can be further expressed in terms of Khatri-Rao product
pseudo-inverse form as 
\begin{align}
  \widetilde{A} = X_{(1)} (C \odot B) (B^{T}B * C^TC)^{\dagger}. 
\end{align}

Here, ${X}_{(1)}$ denotes mode-1 matricization of $\mathcal{X}$. A mode-$n$
matricization flattens or unfolds a $N$-order tensor into a matrix along mode $n$. The
$\odot$ symbol expresses Khatri-Rao product. $(B^{T}B * C^TC)^{\dagger}$ 
computes the pseudo-inverse of the $R \times R $ matrix generated by the dense-dense matrix 
multiplication between $B^{T}B$ and  $C^TC$. Algorithm \ref{alg:cpd}
illustrates the CPD-ALS algorithm; lines \ref{upA}, \ref{upB}, and \ref{upC}
correspond to the update expression for the factor matrices A, B, and C
respectively. 
 
 \begin{algorithm}
   \SetKwInOut{Input}{Input}
   \SetKwInOut{Output}{Output}
   \SetKw{To} {\textbf{to}}
   \SetKw{return} {\textbf{return}}
   \Input{$X \in \mathbb{R}^{I \times J \times K} $} 
   \Output{$A \in \mathbb{R}^{I \times R}, B \in \mathbb{R}^{J \times R}, C \in \mathbb{R}^{K \times R} $\\
    }
  \For{iter = 1 \To outer iters or convergence} {
    A $\leftarrow$ X$_{(1)}$ (C $\odot$ B) (B$^{T}$B * C$^T$C)$^{\dagger}$ \label{upA} \\
    B $\leftarrow$ X$_{(2)}$ (C $\odot$ A) (A$^{T}$A * C$^T$C)$^{\dagger}$ \label{upB}\\
    C $\leftarrow$ X$_{(3)}$ (B $\odot$ A) (A$^{T}$A * B$^T$B)$^{\dagger}$ \label{upC}\\
    normalize columns of A, B, C and store in $\lambda$ \\
  } 
  \return A, B, C
  \caption{CPD-ALS for third-order tensors}
  \label{alg:cpd}
\end{algorithm}

\subsection{Sparse MTTKRP}

A key performance bottleneck of a sparse CPD iteration is the
matricized-tensor times Khatri-Rao product (MTTKRP). A mode-n MTTKRP operates
on all factor matrices except mode-$n$.  For example, the mode-1 MTTKRP of
$\mathcal{X}$ computes
\begin{align}
 Y = {X}_{(1)}(B  \odot C) \label{eq:giga}
\end{align}

The output of $(B  \odot C)$ is a dense $JK \times R$ matrix, which can require
more memory than the tensor itself. Kang et al. proposed a parallel CPD
algorithm, GigaTensor \cite{kang2012gigatensor}, which avoids the $(B  \odot
C)$ operations by separately computing Hadamard products between $B(:,r)$ and
$X_{(1)}$, and $C(:,r)$ and  $X_{(1)}$. As shown in Equation \ref{eq:hmp},
such Hadamard products are computed at each nonzero location of $X_{(1)}$.
\begin{align}
 N_{(i,y)} &= {X}_{(i,y)}B(y\%J,r)C(y/J, r) \label{eq:hmp} 
\end{align}

In equation \ref{eq:hmp}, $N_{(i,y)}$ and ${X}_{(i,y)}$ are the nonzero
entries of the output and input tensors, respectively. $N$ and $X$ have the same
sparsity pattern. Equation \ref{eq:rowSPLATT} reformulates Equation
\ref{eq:giga} for a row of $Y$. Smith et al.  \cite{smith2015tensor} further
optimize the computation by factoring out $C$ as shown in
\ref{eq:BCsplit}.

\begin{align}
 Y(i,:) &= \sum_{z=0}^{JK} X_{(1)}(i,z) (B(z\%J,:) * C(z/J,:)) \label{eq:rowSPLATT} \\
        &= \sum_{k=0}^{K} \sum_{j=0}^{J} X(i,j,k)(B(j,:) * C(k,:)) \label{eq:BCtogether} \\
        &= \sum_{k=0}^{K} C(k,:) * \sum_{j=0}^{J} X(i,j,k)(B(j,:) \label{eq:BCsplit}
\end{align}

This optimization saves $R(J-1)$ multiplications and memory accesses for every
$\mathcal{X}_{(i,:,k)}$ fiber. Similar formulations can be derived for mode-2 and mode-3
MTTKRP for $\mathcal{X}$.

\section{Sparse Tensor Formats}
This section illustrates two families of state-of-the-art sparse tensor formats.

\subsection{COO based formats}
\label{subsec:coo}

COO (coordinate) is a straightforward approach to store a tensor.
For every nonzero entry in the tensor, COO stores a
tuple comprising its indices and its value. A third-order tensor $\mathcal{X}$ will
therefore have $M$ $(i,j,k,\mathit{val})$ items in COO format. For an order-N tensor, the indices
would require $4\times N\times M$ bytes of storage, assuming each index
is a 4-byte integer. 
We note that the analysis of storage requirements for different sparse tensor representations
in this paper only includes that required for the indices and excludes the space
required for the numerical values of the non-zero elements, which will depend on
the precision.
MTTKRP with third-order tensors using the COO
representation is shown in Algorithm \ref{algo:COO}.  Each nonzero
$X_{(i,j,k)}$ multiplies the $j^{th}$ row of B and $k^{th}$ row of C, and
updates the $i^{th}$ row of $A$, requiring $N\times R$ operations.
The total number of operations is $N\times M\times R$. In summary, for a third-order tensor,
\begin{equation*}
\small
\begin{split}
\text{COO}_{\mathit{operations}} &= 3MR\\
\text{COO}_{\mathit{storage}} &= 4\times3M \text{ bytes} 
\end{split}
\end{equation*}

The simplicity and amenability to parallelize over the nonzeros make
COO a popular format \cite{hicoo-li},
\cite{liu2017unified}, \cite{parti}. However, a disadvantage of the parallel
COO format is the requirement of atomic operations to update the output matrix.
Recently developed formats such as FCOO \cite{liu2017unified} and HiCOO \cite{hicoo-li} avoid
atomic operations -- FCOO uses a segmented scan to
avoid the write conflicts, and HiCOO splits a tensor into 
multi-dimensional superblocks and uses a privatization method to eliminate the atomics.

\begin{algorithm}[h!]
\footnotesize
    \SetKwInOut{Input}{Input}
    \SetKwInOut{Output}{Output}
    \SetKw{To} {\textbf{to}}
    \SetKw{Step} {\textbf{step}}
    \SetKw{In} {\textbf{in}}
    \SetKw{return} {\textbf{return}}

    \Input{$indI$[M], $indJ$[M], $indK$[M], $vals$[M], \\
    dense matrices $B$[J][R], $C$[K][R]} 
    \Output{dense matrix $Y$[I][R]} 

    \BlankLine
    \For{z = 0 \To M} {
        i = $indI$[$z$] \\
        j = $indJ$[$z$] \\
        k = $indK$[$z$] \\
     	\For{r = 0 \To R} {
    		$Y[i][r] += vals[z] * B[j][r] * C[k][r]$ \\
    		}
    	}
    	\return $Y$
\caption{COO-MTTKRP for third-order tensors \cite{bader2012matlab}}
\label{algo:COO}
\end{algorithm}

\begin{figure}[htb]
\begin{minipage}{.5\textwidth}
	\centering
     \includegraphics[width=8cm]{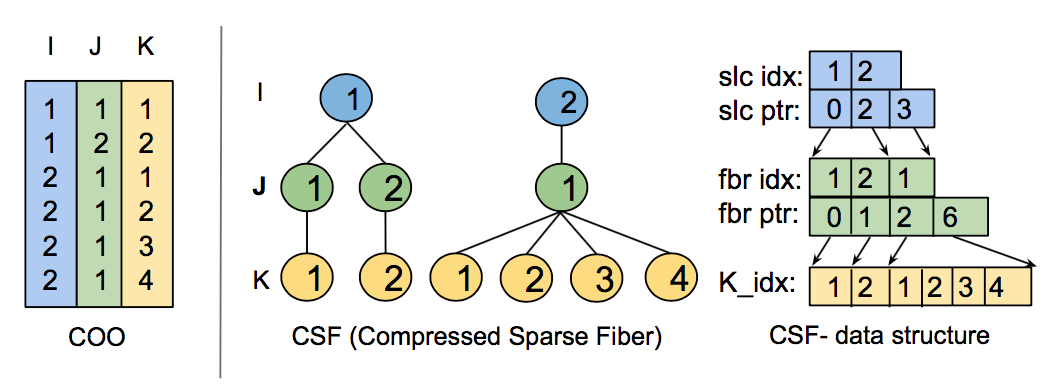}
     \caption{COO and CSF data structures}
     \label{fig:COOnCSF}  
\end{minipage}%
\end{figure}

\subsection{CSR based formats}
\label{subsec:csr}

The most popular format for a sparse matrix is compressed sparse row (CSR).
CSR optimizes COO by grouping nonzeros by row and avoiding explicit storage of row indices, storing only a pointer to indicate the start position of a row.
A hyper-sparse matrix, where its number of nonzeros is much smaller than its
number of rows, contains many empty rows. To better compress
these matrices, Buluc et al.  \cite{buluc2008representation} proposed the doubly compressed sparse row (DCSR) format, where only the pointers for non-empty rows are stored along
with their indices. An extension of DCSR to
tensors is the compressed sparse fiber (CSF) format proposed by Smith et al.
\cite{smith2015splatt}. Figure \ref{fig:COOnCSF} shows the CSF data structure
for a third-order tensor. For a tensor with $S$ slices and $F$ fibers, the storage
requirement for the indices is $4\times (2S + 2F + M)$ bytes, assuming
each index is a 4-byte integer -- $S$
slice pointers, $S$ slice indices, $F$ fiber pointers, $F$ fiber indices, and
$M$ entries to
store the indices for the last dimension (e.g., $K$). 
As shown in \Cref{algo:CSF}, \Cref{line:itSlc} iterates over the slices and \Cref{line:itFbr} goes through the fibers of those slices. In \Cref{line:CSF}, the value of each nonzero of the fibers is multiplied with the corresponding row of $C$, followed by an accumulation in a temporary array $(tmp[])$. The accumulated value is then multiplied by the corresponding row of $B$ (\Cref{line:mulB}) and written back to the row of $A$.
\begin{algorithm}[]
\footnotesize
    \SetKwInOut{Input}{Input}
    \SetKwInOut{Output}{Output}
    \SetKw{To} {\textbf{to}}
    \SetKw{Step} {\textbf{step}}
    \SetKw{In} {\textbf{in}}
    \SetKw{return} {\textbf{return}}

    \Input{$slicePtr$[S], $sliceInds$[S], $fiberPtr$[F], $fiberInds$[F], $indK$[M], $vals$[M],\\
    dense matrices $B$[J][R], $C$[K][R]} 
    \Output{dense matrix $Y$[I][R]} 

    \BlankLine
    \Comment{slices $\mathcal{X}(i,:,:)$ } \\
    \For{slice = 0 \To S  \label{line:itSlc}} 
    {
    	i = $sliceInds$[slice] \\
        \Comment{fibers of each slice} \\
        \For{fiber = $slicePtr$[slice] \To $slicePtr$[slice + 1]  \label{line:itFbr}} 
        {
    	   j = $fiberInds$[fiber] \\
           \Comment{nonzeros of each fiber} \\
            \For{z = $fiberPtr$[fiber] \To $fiberPtr$[fiber + 1]  \label{line:itnnz}} 
            {
    	       k = $indK$[$z$] \\
     	       \For{r = 0 \To R} {
    		        $tmp[r] += vals[z] *  C[k][r]$ \label{line:CSF}\\
    		    }
    	   }
            \For{r = 0 \To R} {
                $Y[i][r] += tmp[r] * B[j][r]$ \label{line:mulB}
            }
        }
    }
	\return $Y$
\caption{CSF-MTTKRP for third-order tensors \cite{smith2015splatt}}
\label{algo:CSF}
\end{algorithm}

CSF has potential to reduce storage requirements, as well
as the required operations depending on the structure of the input tensor. Sparse tensors
often have structures where $S \ll M$ and/or $F \ll M$. For such tensors, the
storage requirement is approximately $4\times M$ bytes. However, for 
tensors where $S \approx F \approx M$, the required storage becomes $4\times
5M$ bytes. The same conclusion can be drawn for operation counts as well. When 
$S \approx F \approx M$, the number of operations performed is $4*M*R$, as the
nonzeros in a fiber are locally reduced before multiplying with the row of
$C$ (Equation \ref{eq:BCsplit}). If $F \ll M$, factored $C$ can save $R(J-1)$
multiplications per fiber, and the operation count can be reduced to $3\times M\times R$. If
$S \ll M$ as well, the operations performed are further reduced to $2\times M\times R$ by
avoiding the addition operation at each row of $M$. The operation and storage count for a third-order tensor
are given by:

\begin{equation*}
\label{eq:CSF}
\small
\begin{split}
\text{CSF}_{\mathit{operations}} = 2(S+M)R &=  
\begin{cases}
2MR, & S, F \ll M \\
4MR, & S, F \approx M \\
\end{cases} \\
\text{CSF}_{\mathit{storage}} = 4 (2S + 2F + M) &= 
\begin{cases}
4 \times 1 M, & S, F \ll M \\
4 \times 5 M, & S, F \approx M \\
\end{cases}
\text{ bytes} 
\end{split}
\end{equation*}

In contrast, regardless of the structure of the tensor, the operation count for COO is $3\times M\times R$
and storage requirement is $4\times3M \text{ bytes}$.

\label{sec:formats}

\section{B-CSF: A Balanced CSF for GPUs}
\label{sec:balanced-sparse-tensor}

\begin{figure*}[htbp]
\centering
\begin{minipage}[t]{0.9\linewidth}
    \includegraphics[width=\linewidth]{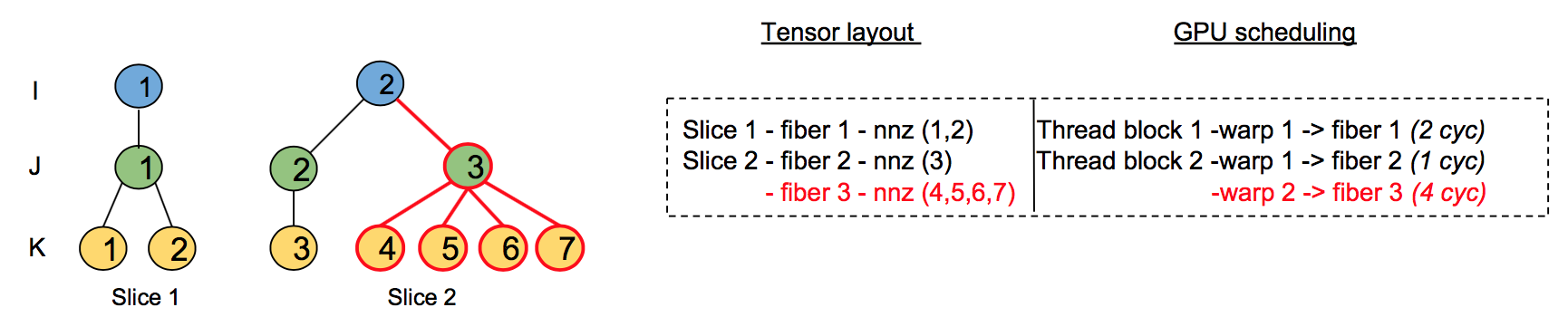}
    \subcaption{Original Tensor causing inter-warp and inter-thread block load imbalance. Required cycle 4. }
    \label{ldBal1}
\end{minipage}%
    \hfill%
\begin{minipage}[t]{0.9\linewidth}
    \includegraphics[width=\linewidth]{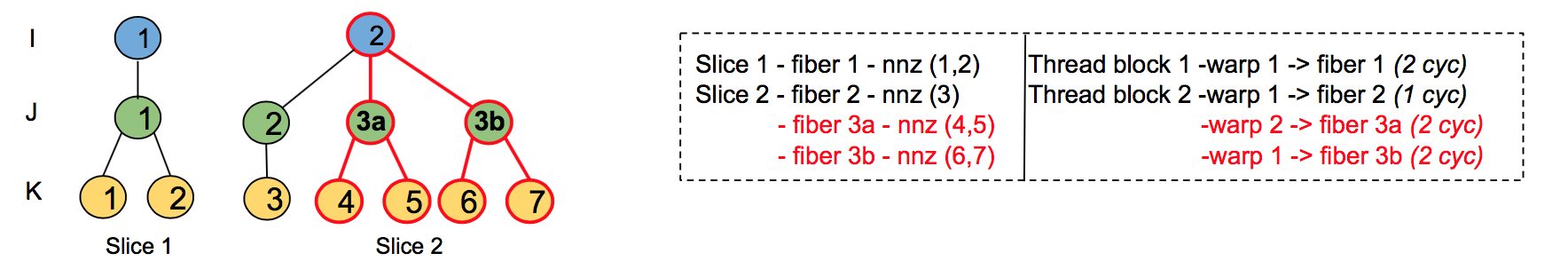}
    \subcaption{After splitting fibers to address inter-warp load imbalance. Required cycle 3.}
    \label{ldBal2}
\end{minipage}
    \hfill%
\begin{minipage}[t]{0.9\linewidth}
    \includegraphics[width=\linewidth]{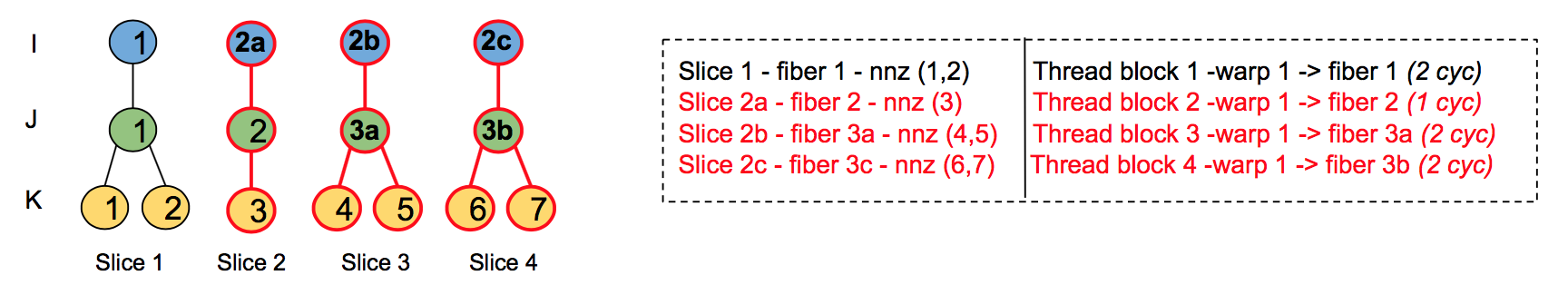}
    \subcaption{After splitting slices along with fibers to address inter-thread block load imbalance. Required cycle 2.}
    \label{ldBal3}
\end{minipage}  
\caption{Construction phases of B-CSF (balanced CSF). Assume one cycle is required to process one nonzero. }
\label{fig:ldbal}
\end{figure*}

Using CSF as a storage format can provide opportunities to reduce operations
and random memory accesses. However, extending CSF to massively parallel
architectures like GPU poses challenges. MTTKRP requires a
reduction across all the nonzeros of the fibers of a slice. On CPUs, SPLATT
uses the CSF data structure, and assigns one thread to process an entire slice in
order to avoid atomics and locks. The CPU thread can be analogous to a warp or
thread block in GPUs\footnote{We use CUDA terminology in this paper.}.  Assume
that a slice is processed by a thread block, fibers within a slice are
distributed across warps in the thread block, and the nonzeros in a fiber across
the threads within a warp.
Like matrices,
real-world tensors tend to follow a power-law distribution. This might
create extensive load imbalance across warps and/or thread blocks. In 
the tested  datasets, the standard deviation of the work distribution across
slices can be as large as 2M (freebase-music, mode 3). A third-order tensor can incur
load imbalance at two levels: (1) the number of nonzeros in the slices
assigned to different thread blocks can vary significantly, leading to
inter-thread block load imbalance; and (2) the number of nonzeros in the
fibers assigned to different warps within a thread block can vary
significantly, creating inter-warp load imbalance.

We collect relevant metrics using the NVIDIA profiler, nvprof \cite{metrics} to
quantify the impact of such load imbalance on performance. We particularly 
focus on two metrics: \textit{sm\_efficiency}, defined by NVIDIA as the percentage of time when at
least one warp is active on a streaming multiprocessor (SM), and 
\textit{achieved\_occupancy}, defined as the ratio of the average active
warps per active cycle to the maximum number of warps supported on an SM. For a number
of tensors used in recent publications, Table \ref{tbl:ldimb} tabulates these metrics, along with  
the achieved performance (GFLOPs) on an Nvidia  P100 GPU. The tensors that exhibit the lowest performance (nell2
and darpa) and high load imbalance, also exhibit high standard deviation in
number of nonzeros per fiber and slices. To address this load-balancing problem, we  
construct a well-balanced tensor.  We term our GPU implementation of CSF as B-CSF,
symbolizing Balanced-CSF.
\begin{table}[htb]
\center
\caption{Performance (GFLOPs) and load imbalance data for Nvidia Tesla P100 }
\label{tbl:ldimb}
\begin{tabular}{ccccccc}
\hline
Tensors   & GFLOPs & \begin{tabular}[c]{@{}c@{}}achv \\ occp. \\ \%\end{tabular} & \begin{tabular}[c]{@{}c@{}}sm\\ effic.\\ \%\end{tabular} & \multicolumn{1}{c}{\begin{tabular}[c]{@{}c@{}}L2 hit\\ rate\end{tabular}} & \begin{tabular}[c]{@{}c@{}}stdev \\ \#nnz\\  per slc\end{tabular} & \begin{tabular}[c]{@{}c@{}}stdev \\ \#nnz\\ per fbr\end{tabular}  \\

\hline

deli         & 90  & 60    & 70   & 62 & 1,011   & 4    \\
nell1       & 33  & 32    & 44    & 20 & 1,314   & 61   \\
\textbf{nell2}      & \textbf{13}   & \textbf{10}    & \textbf{26} & \textbf{83}    & \textbf{27,983}  & \textbf{203} \\
flick-3d   & 46   & 53    & 37  & 67  & 1,851   & 4    \\
fr\_m    & 18   & 65     & 27   & 28  & 105     & 0    \\
fr\_s    & 24   & 67     & 34   &28   & 90      & 0    \\
\textbf{darpa}         & \textbf{2}  & \textbf{4}     & \textbf{12}  &\textbf{4}   & \textbf{25,849} & \textbf{8,588} \\                                                   
\hline
\end{tabular}
\end{table}

\subsection{Addressing inter-thread block load imbalance} 
In our initial work distribution strategy, we assumed that a thread block works
independently on a slice. In an extreme scenario, where the entire computation
is dominated by a single heavy slice, only one thread block will be busy
processing that slice, while the other thread blocks that have finished processing
the other slices will remain idle. As a result, the GPU unit will be blocked
until that working thread block is released. To avoid such a scenario, 
we split the heavy slices into multiple sub-slices. We extend the binning idea proposed
by Ashari et al. \cite{ashari2014fast} for SpMV to determine the number of
thread blocks that are assigned to a slice. For example, if thread blocks contain 512 threads, 
a slice with 2048 nonzeros
$(4 \times 512)$  will be processed by 4 such thread blocks. Note that this distribution may
increase the number of atomic operations across the thread blocks assigned to
the same slice. However, the cost of the extra atomic operations is well
tolerated by the increase in concurrency. We term this optimization technique as
slc-split.

\subsection{Addressing inter-warp load imbalance}
The warps within a thread block will process independent fibers according to
our initial work distribution strategy. However, a heavy fiber can take much
longer processing time than the other fibers, stalling idle warps in the
thread block. In Table \ref{tbl:ldimb}, we see the imbalance in work per fiber
in the last column. This motivates node-splitting:
long fibers are split into fiber-segments.
This enables near-equal workload to all the warps in the thread
block. This preprocessing step can be done while constructing the CSF data
structure, thereby avoiding any additional pre-processing time.  We term this optimization technique as
fbr-split.

Figure \ref{fig:ldbal} summarizes the key load imbalance issues that arise
with work distribution, and shows the construction phases of the new balanced
tensor. We start with \Cref{ldBal1}, where slice 2 and fiber 3 of the tensor has uneven work distribution across warps and thread blocks. To address this issue at the warp level, 
fiber 3 is split into two fibers (3a and 3b) as shown in \Cref{ldBal2}. Warp 1 and 2 from thread block 2 cyclically process the nonzeros. If a multiply-add requires one cycle, cycle count is reduced from four to three. \Cref{ldBal3} demonstrates the
slc-split technique to address load imbalance at the thread block level by splitting slice 2 (2a, 2b, and 2c), and reducing cycle count to two.



	 


\section{HB-CSF: A Hybrid B-CSF}

\begin{figure}[]
\begin{minipage}{.5\textwidth}
   \centering
     \includegraphics[width=7cm]{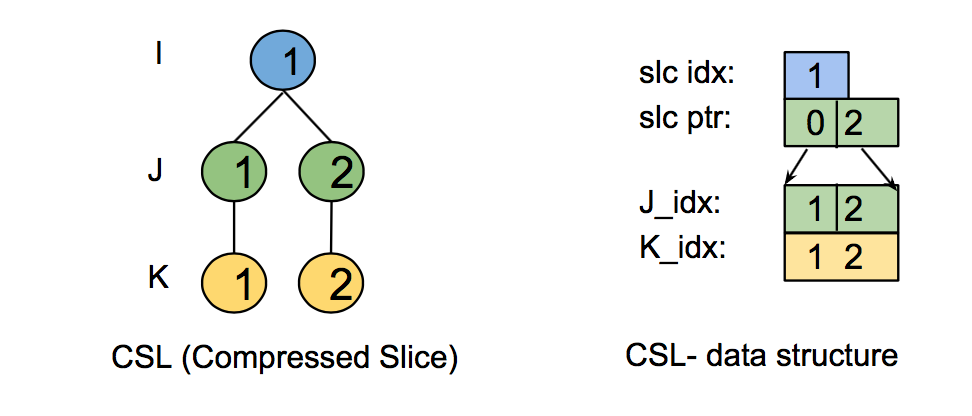}
     \caption{Supported slice type and data structure of CSL}
     \label{fig:CSL}  
\end{minipage}%
\end{figure}

\begin{figure*}[]
    \centering
     \includegraphics[width=18cm]{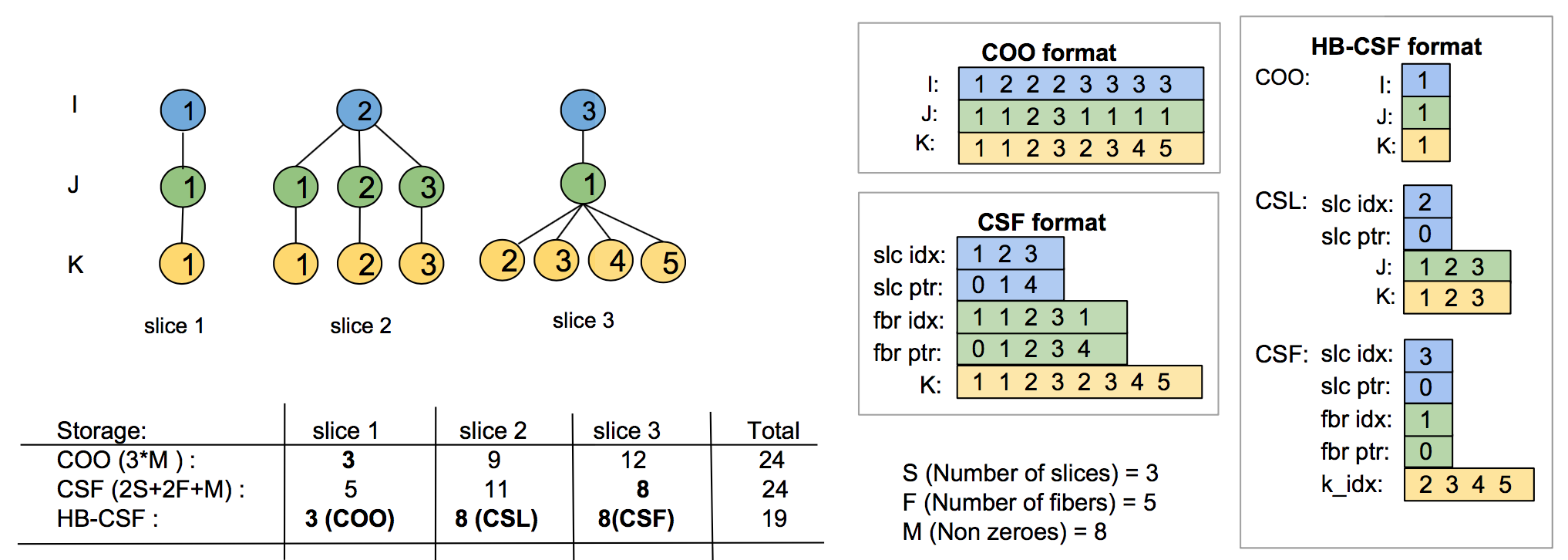}
     \caption{The comparison of a tensor storage using COO, CSF and HB-CSF formats}
     \label{stoFormat}  
\end{figure*} 

The previously discussed splitting techniques addressed the load-imbalance problem posed by heavily populated slices and fibers in tensors.
In this section, we address the opposite problem: inefficiency due to ultra-light slices and fibers.
In many tensors, many fibers may have a single nonzero, and even an entire slice may contain just
a single nonzero element in a single fiber.

\subsection {CSL (Compressed slice)}
If a slice with a single nonzero is stored in a CSF format, its slice and fiber
pointers are redundant. Similarly, when each fiber of a slice has only one
nonzero, fiber pointers become unnecessary. In this case, the slice pointers can directly point
to the nonzeros in the slice and avoid the indirection through a fiber pointer.
We term this technique \textbf{compressed slice} (CSL). As shown in \Cref{fig:COOnCSF}, two slices are presented using CSF format. Now, the slice on the left
is a perfect fit for CSL format, as it has one nonzero at each fiber. In \Cref{fig:CSL}, we show how the CSL data structure can reduce the storage by omitting the fiber pointer (fbr\ ptr) array. This
technique reduces the operation counts as well, as the local reduction across nonzeros of
each fiber is now avoided. \Cref{algo:CSL} demonstrates the steps of an MTTKRP operation using the CSL format. \Cref{line:CSL} iterates through the slices and fetches the nonzero indices and values directly from the slice pointers. Thus, it also avoids the addition operation as in CSF ( \Cref{line:CSF} in \Cref{algo:CSF}).

\begin{algorithm}[h!]
\footnotesize
    \SetKwInOut{Input}{Input}
    \SetKwInOut{Output}{Output}
    \SetKw{To} {\textbf{to}}
    \SetKw{Step} {\textbf{step}}
    \SetKw{In} {\textbf{in}}
    \SetKw{return} {\textbf{return}}

    \Input{$slicePtr$[S], $sliceInds$[S], $indJ$[M], $indK$[M], $vals$[M],\\
    dense matrices $B$[J][R], $C$[K][R]} 
    \Output{dense matrix $Y$[I][R]} 

    \BlankLine
    \Comment{slices $\mathcal{X}(i,:,:)$ } \\
    \For{slice = 0 \To S} 
    {
        i = $sliceInds$[slice] \\

        \Comment{nonzeros of each slice} \\
        \For{z = slicePtr[slice] \To slicePtr[slice + 1] \label{line:CSL}}  {  
            j = $indJ$[$z$] \\
            k = $indK$[$z$] \\
            \For{r = 0 \To R} {
            	$Y(i,r) += vals[z] * B[j][r] * C[k][r]$ \\
        	}
    	}
    }
    \return $Y$
\caption{CSL-MTTKRP for third-order tensors}
\label{algo:CSL}
\end{algorithm} 

\begin{algorithm}[]
\footnotesize
    \SetKwInOut{Input}{Input}
    \SetKwInOut{Output}{Output}
    \SetKw{To} {\textbf{to}}
    \SetKw{Step} {\textbf{step}}
    \SetKw{In} {\textbf{in}}
    \SetKw{return} {\textbf{return}}

    \Input{$~$ $Slices$ $\mathcal{X}(i,:,:)$, $Fibers$ $\mathcal{X}(i,j,:)$}
    \Output{$~$ \textit{HB-CSF}}

    \BlankLine
    \Comment{evaluate slice patterns } \\
    \For{slice = 0 \To S} 
    {
        fiberLenOne = false \\
        
        \For{fiber = slicePtr[slice] \To slicePtr[slice + 1]} 
        {
            nnzFiber = fiberPtr[fiber +1] - fiberPtr[fiber] \\
            \If {nnzFiber $>$ 1 }{
            \textit{fiberLenOne} = false
            }
            nnzSlice += nnzFiber \\
        }
    }
    \Comment{create HB-CSF} \\
    \For{slice = 0 \To S} 
    {
        \If {nnzSlice == 0 }{
            \textit{sliceInCOO} $\cup=$ \textit{slc}
        }
        \ElseIf{fbrLenOne}{ 
            \textit{sliceInCSL} $\cup=$ \textit{slc}
        } 
    	\Else{ 
            \textit{sliceInCSF} $\cup=$ \textit{slc} 
        }
    }   
    \Comment{execute HB-CSF} \\
    \textbf{COO-MTTKRP} \textit{(sliceInCOO)} \\
    \textbf{CSL-MTTKRP} \textit{(sliceInCSL)} \\
    \textbf{CSF-MTTKRP} \textit{(sliceInCSF)} \\
\caption{HB-CSF-MTTKRP for third-order tensors}
\label{algo:HYB}
\end{algorithm} 



The opportunities to save storage and operations depending on the nonzero distribution pattern are incorporated in a new storage format, HB-CSF (Hybrid B-CSF). Tensor slices fall in one of three groups: i) those with a
single nonzero, ii) those with only singleton fibers (i.e., each fiber in the slice  has a single nonzero),
and iii) all other slices.  The nonzeros in these three groups of slices are stored using COO, CSL and CSF
formats,
respectively, to generate the hybrid HB-CSF format. As demonstrated in Algorithm \ref{algo:HYB},
the slices with a single nonzero are added to the \textit{sliceInCOO} array;
slices with one nonzero at each fiber are put in the \textit{sliceInCSL}
array; the remaining slices use the CSF data structure. A combination of these three
formats is shown in Figure \ref{stoFormat}. It has three slices ($S = 3$), five fibers ($F =5$ ), and eight nonzeros($M=8$). A COO representation of the tensor
in this figure requires 24 words for the indices $(3 * M)$. A CSF format will need the same number of words for the indices $(2 * S + 2* F + M )$. However, when the space requirements for an individual slice is computed, the third slice requires four fewer words than COO. Finally, using the HB-CSF format, with slice 1 in COO, slice 2 in CSL and slice 3 in CSF, the storage requirement is 19 words. In summary, for a third-order tensor,

\begin{equation*}
\begin{split}
\text{HB-CSF}_{\mathit{operations}} &= 2MR \sim 3MR \\
\text{HB-CSF}_{\mathit{storage}} &= 4 \times (1M \sim 3M ) \text{ bytes} 
\end{split}
\end{equation*} 

\label{sec:hyb}

\section{Experiments}
\subsection{Experimental setting}

We evaluate our work against state-of-the-art frameworks on both CPU and
GPU platforms. The experimental evaluation on GPUs is done on an NVIDIA Tesla
P100 (Pascal) device. The P100 device contains 56 SMs, with 16 GB global
memory, 4096 KB L2 cache, a peak single-precision performance of 9.3 TFLOPS,
and a peak memory bandwidth of 732 GB/s. The experimental evaluation on CPUs is done on a Dell PowerEdge R730 two-socket servers with Intel Xeon E5-2680
v4. It is an Intel Broadwell platform with 28 cores. The base
frequency is 2.40GHz with 128GB memory, and 35 MB L3 cache. We use 32 bit
unsigned integers to store the indices and 32 bit floats to store the values.
The GPU codes are compiled with NVCC-9.2 and the CPU codes are compiled with
icc-16.0.3. Source code is available at \footnote{https://github.com/isratnisa/B-CSF}.

Most of the datasets used in this work are collected from The Formidable
Repository of Open Sparse Tensors and Tools, FROSTT \cite{frosttdataset}. We use ch-cr
and flick as abbreviations for the chicago-crime and flickr datasets. 
The remaining datasets, like darpa, fr\_m (freebase-music) and 
fr\_ms (freebase-sampled) are from the dataset of HaTen2 \cite{jeon2015haten2}. Table
\ref{dataset} shows the order, number of nonzeros (\#Nonzeros), and the
density (defined as $(\#Nonzeros/(I \times J \times K))$ of the tensors. We
compare the performance of HB-CSF with two recent works on the GPU platform, FCOO \cite{liu2017unified}
and ParTI \cite{parti}. 
As suggested by the authors of FCOO \cite{liu2017unified}, we tune the framework 
for thread block sizes in \{32,
64, 128, 256, 512, 1024\} and threadlen sizes in \{8, 16, 32, 64\}. The ParTI library
provides MTTKRP for both GPU and CPU (OpenMP) platforms. We use the best
parameter configurations suggested by the authors. SPLATT \cite{smith2015splatt}
is the state-of-the-art CPU framework. SPLATT provides a user-defined option
to select the number of CSF representations for the input tensor. For example,
the ONEMODE setting keeps a single CSF representation for one particular mode
(root mode), and performs MTTKRP for all modes using that single representation. 
Except for the
root mode, MTTKRP for other modes is performed via recursion, which causes performance
degradation.  Hence, in this work, we use the most efficient ALLMODE setting and store $N$ CSF
formats to achieve maximum performance. SPLATT also provide an optimization 
flag - tiling to exploit data locality. We evaluate the performance of SPLATT by
both enabling and disabling this option. SPLATT
release version 1.1.0 is used in this work. We also compare performance with
HiCOO \cite{hicoo-li}, developed for CPUs. $R$ is 32 for all the experiments.

\begin{table}[]
\scriptsize
\center
\caption{Sparse tensor datasets}
\label{dataset}
\begin{tabular}{lllll}

Tensors       &order                & Dimensions                    & \#Nonzeros    & Density\\ \hline
deli            &3     & $533K \times 17M \times 2M$   & 140M  & 6.14E-12 \\
nell1           &3          & $3M   \times 2M  \times 25M$  & 144M  & 9.05E-13 \\
nell2           &3            & $12K  \times 9K  \times 29K$  & 77M   & 9.05E-13 \\
flick           &3              & $320K \times 28M \times 2M$   & 113M  &7.80E-12 \\
fr\_m           &3      & $23M  \times 23M \times 166$  & 99M   &1.10E-09 \\
fr\_s           &3    & $39M  \times 39M \times 532$  & 140M  &1.73E-10 \\
darpa           &3             & $22K  \times 22K \times 23M$  & 28M   &2.37E-09 \\ \hline
nips            &4               & $2K \times 3K \times 14K \times 17$          & 3M   &3.85E-04 \\
enron           &4            & $6K \times 6K \times 244K \times 1K$         & 5M   &1.83E-06 \\
ch-cr           &4           & $6K \times 24 \times 77 \times 32$           & 54M  &1.48E-01 \\
flick           &4                 & $320K \times 28M \times 2M \times 731$       & 113M &1.07E-14 \\
uber            &4                   & $183 \times 24 \times 1K \times 2K$          & 3M   &5.37E-10\\

\hline
\end{tabular}
\end{table}

\begin{figure}[t]
    \centering
     \includegraphics[width=9cm]{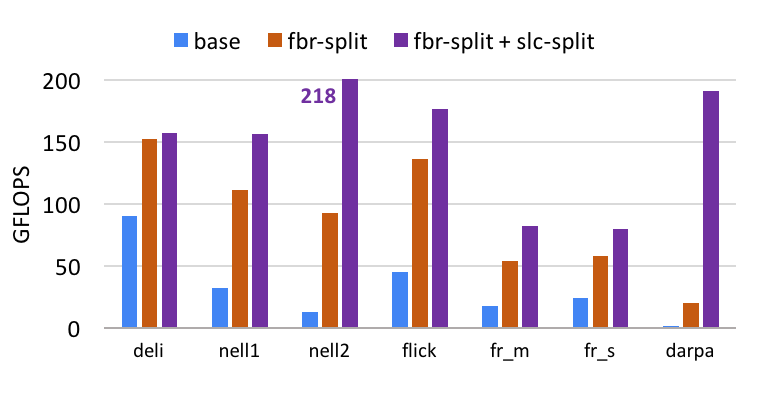}
     \caption{Performance (GFLOPs) with B-CSF in mode 1 using fiber-split and slice-split on P100 GPU}
     \label{fig:nodeSplitting}  
\end{figure}

\begin{figure}[htbp]
    \centering
     \includegraphics[width=9cm]{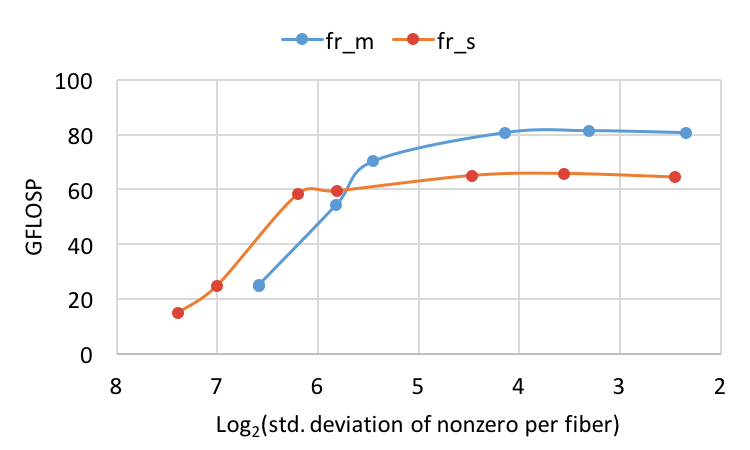}
     \caption{Increase in performance with decrease of standard deviation of nonzeros per fiber: freebase music (fr_m) and freebase sampled (fr_s) datasets in mode 1 }
     \label{fig:devVSGFLOPS}  
\end{figure} 

\subsection{Optimization by node splitting}
In our evaluation, whenever the number of nonzeros in a fiber exceeds a threshold, we split
it into multiple fiber-segments. For example, if a threshold is set at 16, and a fiber
contains 32 nonzeros, it will be split into 2 fiber-segments. We empirically find that
a fiber threshold of 128 provides the best performance. We follow an implicit technique to perform the slice splitting. Instead of
splitting a slice, we increase the number of thread blocks that work on a slice.

\begin{figure}[htbp]
\begin{minipage}[t]{0.48\linewidth}
    \includegraphics[width=\linewidth]{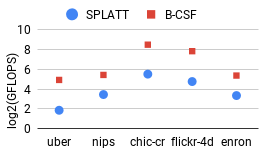}
    \subcaption{shortest mode}
    \label{fig:shorty}
\end{minipage}%
    \hfill%
\begin{minipage}[t]{0.48\linewidth}
    \includegraphics[width=\linewidth]{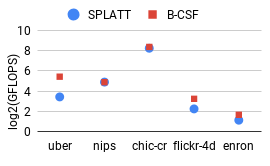}
    \subcaption{longest mode}
    \label{fig:longy}
\end{minipage} 
\caption{Performance (GFLOPs) with CSF from SPLATT and B-CSF}
\label{fig:model}
\end{figure}

The impact of fiber and slice splitting on performance is
shown in Figure \ref{fig:nodeSplitting}. The darpa dataset benefits the most
(a $22 \times$ speedup) from this technique, primarily because it has the highest
standard deviation of work per slice (60,327).  Figure \ref{fig:devVSGFLOPS}
shows the relation between achieved performance (GFLOPs) and standard deviation of nonzeros per fiber (nnzfiber)
on fr\_m and fr\_b tensors. Starting with the original standard deviation of nnzfiber of the tensors, we see that by decreasing the deviation (improving warp level load balance), performance improves.  
The CSF based implementation in the SPLATT library scales poorly on short modes. By using the splitting 
optimization for load balancing, we concurrently resolve the scalability issue.
As shown in Figure \ref{fig:shorty} and \ref{fig:longy}, B-CSF scales well for 
 both the shortest and the longest mode.

\subsection{Impact of using HB-CSF format}

\begin{figure}[htbp]
    \centering
     \includegraphics[width=8cm]{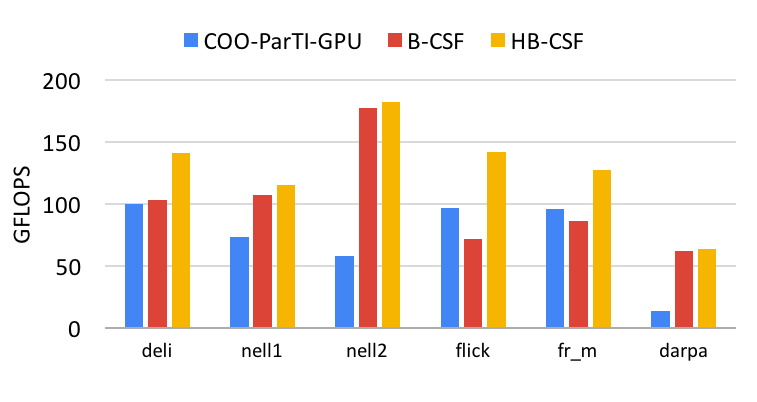}
     \caption{Performance (GFLOPs) using ParTI-COO-GPU, B-CSF and HB-CSF in mode-1 on P100 GPU}
     \label{fig:perfHYB}  
\end{figure}  

Figure \ref{fig:perfHYB} shows that for some cases (e.g., flick-3d
and fm-s) the COO format outperforms the highly optimized B-CSF. In both these
test cases, the average work per slice is 4. Also, in flick-3d, each
fiber has only one nonzero. As explained before in Section
\ref{sec:background}, this results in extra computation for CSF. Since HB-CSF
eliminates this extra computation by carefully choosing between COO, CSF and CSL,
it avoids the overheads, and achieves consistently better performance
for all the cases.

\subsection{Pre-processing}

\begin{figure}[htbp]
    \centering
     \includegraphics[width=8cm]{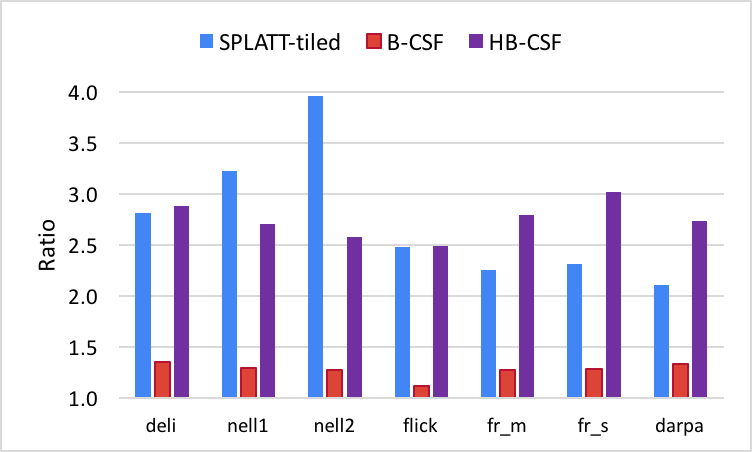}
     \caption{Ratio of pre-processing time compared to SPLATT-nontiled}
     \label{fig:preRatio}  
\end{figure}  

\begin{figure}[htbp]
    \centering
     \includegraphics[width=8cm]{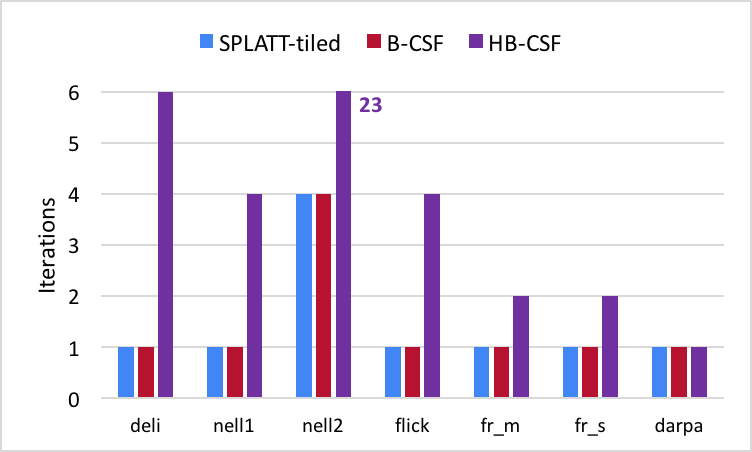}
     \caption{Number of iterations required to outperform SPLATT-nontiled, including pre-processing and execution time}
     \label{fig:iters}  
\end{figure}

We construct the HB-CSF format starting from a CSF data structure. The slices are first
assigned an appropriate format, based on their nonzeros and underlying
structure. A slice is stored as (a) COO if it has one nonzero; (b) CSL if each
fiber has one nonzero; and (c) CSF otherwise. \Cref{fig:preRatio} shows
the preprocessing time for B-CSF, HB-CSF and SPLATT-tiled, normalized to the time
for non-tiled SPLATT. Since CPD is an iterative algorithm and each iteration requires MTTKRP over
all modes, the 
preprocessing cost is amortized over a number of iterations. \Cref{fig:iters} 
shows the number of iteration required for B-CSF and HB-CSF to outperform SPLATT-nontiled, including
execution as well as preprocessing time. B-CSF requires negligible preprocessing time
and generally has comparable performance (\Cref{fig:perfHYB}) to HB-CSF. This makes B-CSF a 
suitable choice for scenarios where the expected number of iterations is low.

\begin{figure*}[htbp]
\begin{minipage}{.5\textwidth}
    \includegraphics[width=9cm]{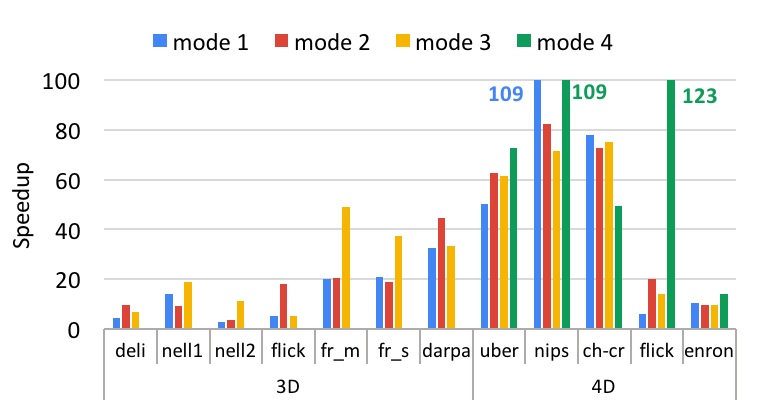}
    \caption{Speedup over SPLATT-CPU-tiled}
    \label{tiling} 
    \end{minipage}%
\hfill
\begin{minipage}{.5\textwidth}
    \includegraphics[width=9cm]{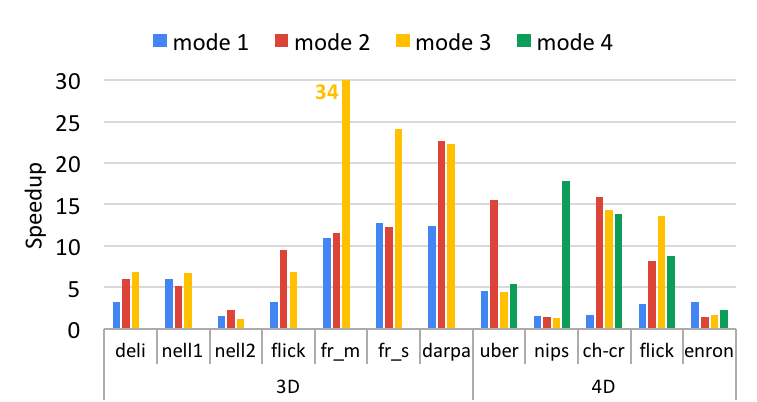}
    \caption{Speedup over SPLATT-CPU-nontiled}
    \label{nontiling} 
\end{minipage}%
\end{figure*}  

\begin{figure*}[htbp]
\begin{minipage}{.5\textwidth}
     \includegraphics[width=9cm]{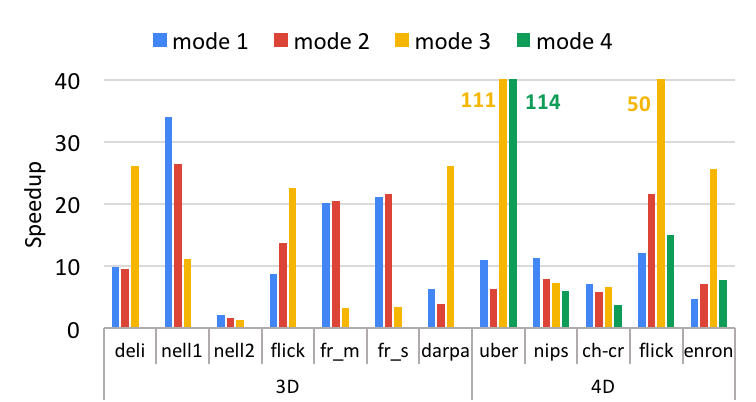}
     \caption{Speedup over HiCOO-CPU }     
     \label{fig:hicoo} 
    \end{minipage}%
\hfill
\begin{minipage}{.5\textwidth}
     \includegraphics[width=9cm]{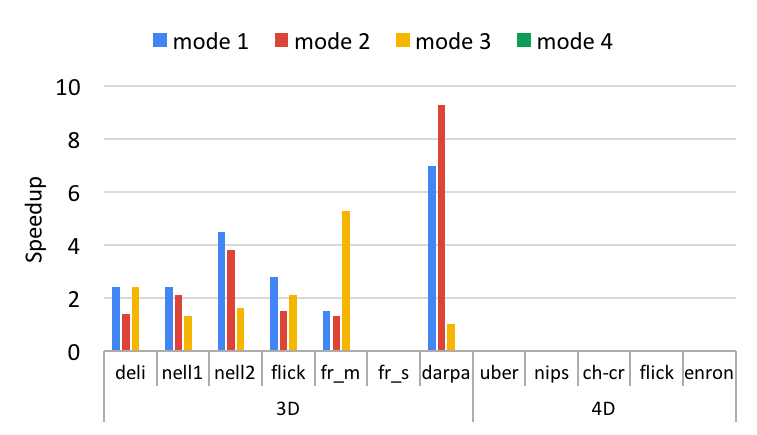}
     \caption{Speedup over ParTI-GPU }
    \label{fig:spdParTIGPU}  
     \end{minipage}%
\end{figure*}

\begin{figure}[htbp]
    \centering
    \includegraphics[width=9cm]{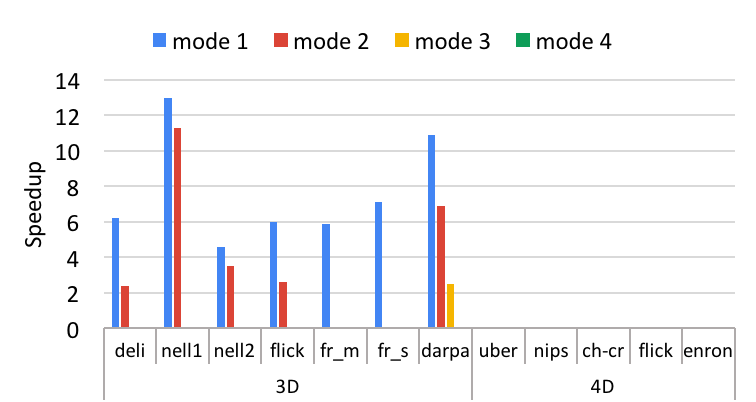}
     \caption{Speedup over FCOO-GPU}
     \label{fig:spdFCOO} 

\end{figure} 

\subsection{Comparison with other frameworks}
In this section, we present performance comparisons with other available MTTKRP
codes.
CPU based benchmarks were run using 28 threads. As shown in \Cref{tiling}, 
HB-CSF outperforms SPLATT (tiling enabled) by $35 \times$ on average. flick-4d achieves the highest 
speedup at its shortest mode (mode 4). This is primarily because the node-splitting technique helps to
achieve higher parallelism and also has enough work (112M nonzeros) to occupy
the entire GPU. However, the optimization flag - tiling can adversely affect the
performance of SPLATT depending on the structure of the tensor. In \Cref{nontiling}, we 
demonstrate the performance with disabling of the tiling option and achieve $9 \times$ speedup on average.
Fig.~\ref{fig:hicoo} compares performance with
HiCOO, 
which is $17 \times$ slower on average.
None of the existing GPU based frameworks, like FCOO and ParTI-GPU, support four or
higher dimensional tensors. That is the reason for the missing bars for the 4D tensors 
in Figure \ref{fig:spdParTIGPU} and \ref{fig:spdFCOO}.
The other missing bars denote the inability of the framework to produce a correct result
across those modes, often due to insufficient memory. 
The FCOO and ParTI-GPU frameworks are outperformed by HB-CSF by $4 \times$
and $3 \times$, respectively. 

\subsection{Storage comparison}

\begin{figure}[htbp]
    \centering
     \includegraphics[width=8cm]{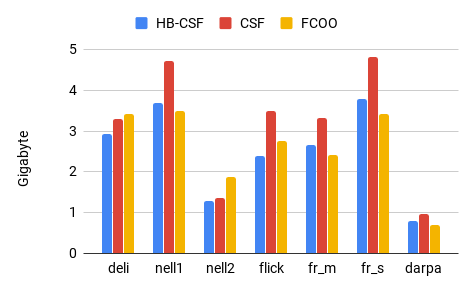}
     \caption{Storage requirements for FCOO, CSF and HB-CSF }
     \label{fig:stoComp}  
\end{figure}

The sparsity pattern of a tensor varies highly across modes e.g., the number of slices, the number of fibers, nonzeros per fiber, nonzeros per slice, etc.  Hence, CSF-like formats with strong mode-orientation ($N$ representations for a $N$-order tensor) have been used by many efforts \cite{choi2018blocking}, \cite{smith2016medium}. Figure \ref{fig:stoComp} compares the storage requirements of formats associated with strong mode orientation, like FCOO and CSF. We account only for the indices, since the numerical values always have the same storage needs in all storage methods. HB-CSF consistently occupies less space than CSF, as it avoids storing redundant pointers. Tensors with repetitive slice and fiber indices get the most benefit from these stores. On the other hand, for the tensors consisting of sparse slices or/and sparse fibers, FCOO require less storage. For each nonzero, FCOO maintains a boolean array to indicate the starting location of the fibers, instead of an integer array like COO. However, the COO portion of HB-CSF can be further optimized in this manner to further reduce storage.

\section{Related Work}

A significant amount of research has been conducted on sparse tensor operations in terms of storage, computation, and performance in this decade. Tensor factorization is gaining significant popularity, like matrix factorization 
\cite{zhou2008large}, \cite{yu2014parallel}, \cite{nisa2017parallel}.
In this section, we briefly discuss prior work on MTTKRP algorithms and sparse tensor formats,
 for both CPU and GPU platforms.

Tensor Toolbox \cite{bader2012matlab} and Tensorlab
\cite{vervliet2016tensorlab} provide COO-MTTKRP implementations, which are computed as a series of sparse tensor-vector products instead of tensor-matrix products. 
Their COO-MTTKRPs uses $3MR$ operations and $M$ words of intermediate storage. 
GigaTensor \cite{kang2012gigatensor} provides a highly
scalable framework, which targets large-scale sparse tensors using the MapReduce
paradigm. 
Its computation requires $5MR$ operations and requires $M + max(J,K)$ words of intermediate storage for a third-order tensor.
Smith et al. \cite{smith2015splatt} have developed the SPLATT library, which includes MTTKRP on
sparse tensors and proposed the CSF storage format, an extension of the Compressed Sparse Row (CSR) format for sparse matrices.
CSF-MTTKRP performs $2R(m+P)$ operations with only $R$ words of intermediate storage.
Their work also applies reordering and cache blocking techniques.
Recently, Choi et al. \cite{choi2018blocking} used two blocking optimizations to further optimize CSF-MTTKRP.
DFacTo \cite{choi2014dfacto} aims to exploit existing high performing 
sparse matrix-vector multiplication (SpMV) routines, and therefore develops an
algorithm to perform an MTTKRP by computing multiple SpMVs. DFacTo computes one column at a
time with two SpMV operations, which requires $2R(M+F)$ operations, where $F$
is the number of non-empty fibers in the corresponding mode. 
The intermediate storage for DFacTo is large since the output of the first SpMV needs to be saved to compute the next one.
A new format called Hierarchical COOrdinate (HiCOO) was recently proposed by Li et al. \cite{hicoo-li}, which is derived from the COO format.
HiCOO compresses tensor indices in units of sparse tensor blocks, thereby saving storage and improving data locality.
HiCOO-MTTKRP is efficient especially for irregularly-shaped tensors and tensors with blocks of relatively high density of non-zeros.
Other research \cite{Blanco:2018,smith2017knl,Li:2017:CPD,Kaya:2018:cpd} has targeted improvement of MTTKRP performance on different platforms, such as the Intel Xeon Phi Knights Landing manycore processor, or considered a sequence of MTTKRPs as a whole to enhance intermediate data reuse.


We are aware of two existing implementations of sparse MTTKRP
for GPUs: 1) ParTI! by Li et al. \cite{li2016optimizing,parti}, and 2) MTTKRP based on Flagged COO (F-COO) format by Liu et
al. \cite{liu2017unified}. 
ParTI! stores the input tensor in COO format and parallelizes over nonzeros. 
It performs an atomic add when combining nonzero products to the same data. 
MTTKRP based on the F-COO format also processes nonzero elements in parallel but uses a parallel scan algorithm
to reduce write conflicts at the same location. 
The F-COO format keeps two boolean arrays to indicate any changes in slice/fiber indices and between CUDA threads, instead of storing the slice/fiber's actual indices.
We have presented experimental comparisons with both these sparse MTTKRP algorithms and demonstrated
significant speedup.

\section{Conclusion}



In this paper, we have addressed the problem of developing an efficient GPU
implementation of the important sparse MTTKRP kernel. We have addressed the
fundamental challenge of enabling both efficient data access and good load-balancing
at multiple levels of GPU parallelism by devising the new HB-CSF
representation for sparse tensors. The sparse MTTKRP implementation using HB-CSF
was demonstrated to be considerably faster than existing sparse MTTKRP
implementations for CPUs or GPUs.

The optimizations incorporated into the HB-CSF based sparse MTTKRP implementation are 
complementary to other optimizations recently proposed
for COO- or CSF-based algorithms, e.g., 3D blocking, various reordering methods (Z-order sorting, graph and hypergraph partitioning). 
Future work will explore integration of some of these complementary strategies
to further improve performance of sparse MTTKRP.

\label{sec:conclusion}


\section*{Acknowledgments}
We thank the reviewers for the valuable feedback and the Ohio Supercomputer Center for use of their GPU resources. This work was supported in part by the Defense Advanced Research Projects Agency (DARPA) under Contract D16PC00183, and the National Science Foundation (NSF) through awards 1404995, 1513120, and 1629548. This research was also partially funded by the US Department of Energy, Office for Advanced Scientific Computing (ASCR) under Award No. 66150: "CENATE:  The Center for Advanced Technology Evaluation". Pacific Northwest National Laboratory (PNNL) is a multiprogram national laboratory operated for DOE by Battelle Memorial Institute under Contract DE-AC05-76RL01830. Also contributing to this research, are funds from the Defense Advanced Research Projects Agency (DARPA) contract FA8750-18-2-0108, under the DARPA MTO Software Defined Hardware program, and the Laboratory Directed Research and Development program at Sandia National Laboratories under contract DE-NA-0003525. Disclaimer: The views, opinions, and/or findings contained in this document are those solely of the author(s) and should not be interpreted as representing the official views or policies of any of its funding sources.

\bibliographystyle{IEEEtran}
\bibliography{references} 

\end{document}